\documentclass{iopart}   
\usepackage{iopams}
\usepackage{epsf}
\usepackage{times}
\eqnobysec

\begin{document}


\markboth{M.S. Bruz\'on, P.G. Est\'evez, M.L. Gandarias and J. Prada} {$1+1$ spectral problems arising from
the Manakov-Santini system}

\title[$1+1$ spectral problems arising from
the Manakov-Santini system]{$1+1$ spectral problems arising from the Manakov-Santini system}

\author{M. S. Bruz\'on\dag, P. G. Est\'evez\ddag, M. L. Gandarias \dag\ and J. Prada $\clubsuit$}

\address{\dag\ Departamento de Matem\'aticas,
Universidad de C\'adiz, Spain}%

\address{\ddag\ Departamento de F\'{\i}sica Fundamental, \'Area de F\'{\i}sica
Te\'{o}rica, Universidad de Salamanca,  Spain}
\address{$\clubsuit$\ Departamento de Matem\'aticas,  Universidad de Salamanca,  Spain}

\begin{abstract}
This paper deals with the spectral problem of the Manakov Santini system. The point Lie symmetries of the Lax
pair have been identified. Several similarity reductions arise from these symmetries. An important benefit of
our procedure is that the study of the Lax pair instead of the partial differential equations yields the
reductions of the eigenfunctions and also the spectral parameter. Therefore, we have obtained five
interesting spectral problems in $1+1$ dimensions.
\end{abstract}

{keywords: Lie symmetries; similarity reductions; spectral problem}


\section{Introduction}

One of the most powerful instruments to study and/or solve a given differential equation is the
identification of the Lie point symmetries of the equation \cite{ibragimov}.  A standard method for finding
solutions of a partial differential equation  (PDE) is that of reduction by using Lie symmetries: each Lie
symmetry allows a reduction of the PDE to a new equation with the number of independent variables reduced by
one.  Classical \cite{ibragimov}, \cite{stephani} and non-classical \cite{bluco}, \cite{olver} Lie symmetries
are the usual way used to identify the reductions. The similarity reductions of PDEs obtained through the
calculation of their Lie symmetries is a standard procedure that has been successfully applied in the
scientific literature for many decades. The connection between these method and/or other methods for
obtaining similarity reductions has also been extensively discussed \cite{estevez93}, \cite{estevez95}.

As  is well known, one of the best proofs of the integrability of a PDE is the existence of a Lax pair, which
requires the introduction of a new dependent field (the eigenfunction) and a constant (the spectral
parameter), which can be also interpreted as a new independent variable such that only the eigenfunction
depends on it.

It is natural to deal with the problem of identifying reductions of the Lax Pair instead of those of the
equations \cite{leg}. The obvious benefit is that, in this case, we know how the eigenfunction and spectral
parameter will reduce \cite{egp05}. This is by no means a trivial question, as we can see in the example that
we are interested here. This example is  the Manakov- Santini system \cite{santini}, which reads:
\begin{eqnarray}
&&u_{xt}+u_{yy}+(uu_x)_x+v_xu_{xy}-v_yu_{xx}=0 \nonumber\\
&&v_{xt}+v_{yy}+uv_{xx}+v_xv_{xy}-v_yv_{xx}=0, \label{1.1}
\end{eqnarray}
whith the Lax pair
\begin{eqnarray}
\psi_y&=&-\left(\lambda+v_x\right)\psi_x+u_x\psi_{\lambda}\nonumber\\
\psi_t&=&-\left(\lambda^2+\lambda v_x +u - v_y\right)\psi_x+\left(\lambda
u_x-u_y\right)\psi_{\lambda},\label{1.2}
\end{eqnarray}
where $u=u(x,y,t)$, $v=v(x,y,t)$, $\psi=\psi(x,y,t,\lambda)$.

Equation (\ref{1.1}) is a member of the Manakov- Santini hierarchy \cite{santini} and it is well known that
it has several interesting reductions \cite{bodganov},\cite{chang},\cite{pavlov}.

According to our previous statement, here we shall address with the Lie point symmetries of the Lax pair
(\ref{1.2}), where $u, v, \psi$
 are the dependent variables and $x,y,t,\lambda$ the independent ones.

The plan of the paper is the following
\begin{itemize}
\item Calculation of the Lie symmetries of the Lax pair (\ref{1.2}) is dealt with in  section 2.
\item The five different reduced spectral problems appear in section 3. The equations obtained from these
spectral problems are also identified. Two of them are linear equations than can be integrated. The other
three systems include equations such as the Monge-Ampere and Modified Hunter-Saxton equations.
\item We close with a section of conclusions.
\item Some necessary but tedious expressions are listed in the appendix.
\end{itemize}

\section{Lie point symmetries of the  spectral problem}
Here, we are interested  in the Lie symmetries of the Lax pair. Actually, the symmetries of equations
\ref{1.1} are interesting in themselves, but we  also wish to know
 how \textbf{the eigenfunction and the spectral parameter transform under the action of a Lie symmetry}.
  More precisely, we wish to know what these fields look like under the reduction associated with each symmetry.
This is why we shall proceed to write the infinitesimal Lie point transformation of the variables
   and fields that appear in the spectral problem.
  The benefits of such a procedure have been shown in \cite{leg} and \cite{egp05}.

   In the present case, it is important to note that the
   spectral parameter appears as an independent variable.

   The infinitesimal form of the Lie point symmetry that we are considering is:

\begin{eqnarray}
 x' &=& x + \varepsilon\,\xi_1(x,y,t,\lambda,\psi,u,v) + O(\varepsilon^2)\nonumber\\
y' &=&y + \varepsilon\,\xi_2(x,y,t,\lambda,\psi,u,v) + O(\varepsilon^2)\nonumber\\
t'  &=& t + \varepsilon\,\xi_3(x,y,t,\lambda,\psi,u,v) + O(\varepsilon^2)\label{2.1}\\
\lambda'  &=& \lambda + \varepsilon\,\xi_4(x,y,t,\lambda,\psi,u,v) + O(\varepsilon^2)\nonumber
\\
u'  &=& u + \varepsilon\,\phi_1(x,y,t,\lambda,\psi,u,v) + O(\varepsilon^2)\nonumber
\\v'  &=& v+ \varepsilon\,\phi_2(x,y,t,\lambda,\psi,u,v) + O(\varepsilon^2)\nonumber\\
\psi'  &=& \psi + \varepsilon\,\phi_3(x,y,t,\lambda,\psi,u,v) + O(\varepsilon^2),\nonumber
\end{eqnarray} where
$\epsilon$ is the group parameter. The associated Lie algebra of infinitesimal symmetries is the set of
vector fields of the form:
\begin{equation} X = \xi_1\frac {\partial}{\partial x} + \xi_2\frac {\partial}{\partial y}+\xi_3\frac
{\partial}{\partial t}+\xi_4\frac {\partial}{\partial \lambda}+\phi_1\frac {\partial}{\partial u}+\phi_2\frac
{\partial}{\partial v}+\phi_3\frac {\partial}{\partial \psi}.\label{2.2}\end{equation}

 We  also need to know
how the derivatives of the fields transform under the Lie symmetry, which means that we have  to introduce
the ``prolongations" of the action of the group  to the different derivatives that appear in (\ref{1.2}).
Exactly how to calculate the prolongations is a very well known procedure whose technical details can be
found in \cite{bluco}, \cite{stephani}.

 It is therefore necessary  that the Lie  transformation should leave (\ref{1.2}) invariant.
 This yields an overdetermined  system of equations for the
infinitesimals $\xi _1(x,y,t,\lambda,\psi,u,v)$, $\xi_2(x,y,t,\lambda,\psi,u,v)$, $
\xi_3(x,y,t,\lambda,\psi,u,v)$, $\xi_4(x,y,t,\lambda,\psi,u,v)$, $\phi_1(x,y,t,\lambda,\psi,u,v)$,
$\phi_2(x,y,t,\lambda,\psi,u,v),$ and $\phi_3(x,y,t,\lambda,\psi,u,v),$.

Below is a summary  of the  \textbf{classical Lie method} \cite{stephani} of finding Lie symmetries
\begin{itemize}
\item  Calculation of the prolongations of the derivatives of the fields that appear in (\ref{1.2})
\item  Substitution of the transformed fields (\ref{2.1}) and their derivatives in (\ref{1.2}).
\item  Set  all the coefficients in $\epsilon$ at $0$.
\item Substitution of the prolongations.
\item  $\psi_y$ and $\psi_{\lambda}$ can be substituted by using (\ref{1.2}).
\item  The system of equations for the infinitesimals can be obtained by setting  each coefficient in
the different remaining derivatives of the fields at zero.
\end{itemize}
From the technical point of view,  calculation of the determining equations can be performed by using
computer packages such as MACSYMA or MAPLE. We have used both independently to determine the equations and
solve them. The result is the following set of symmetries:

\begin{eqnarray}
 \xi_1 &=&-\frac{1}{2}\left(2\alpha-\tau_t\right)_t y^2+\beta y+\left(2\alpha-
 \tau_t\right)x+\gamma\nonumber\\
 \xi_2 &=&\alpha y +\delta\nonumber\\
 \xi_3&=& \tau\nonumber\\
  \xi_4 &=&\left(\alpha-\tau_t\right)_t y+\left(\alpha-\tau_t\right)\lambda-\beta-
  \delta_t\nonumber\\
  \phi_1 &=& 2\left(\alpha-\tau_t\right)u+\frac{1}{2}\left(\alpha-\tau_t\right)_{tt}y^2-\left(\beta+\delta_t\right)_ty-
  \left(\alpha-\tau_t\right)_tx+\theta\label{2.3}\\
   \phi_2&=& \left(3\alpha-2\tau_t\right)v+\left(\frac{\alpha}{2}-\frac{\tau_t}{3}\right)_{tt}y^3-
   \left(\beta+\frac{\delta_t}{2}\right)_ty^2\nonumber\\&&+\left[\left(2\tau_t-3\alpha\right)_tx+\theta-\gamma_t\right]y
   +\left(2\beta+\delta_t\right)x+\sigma\nonumber\\
    \phi_3&=&\Omega(\psi),\nonumber
\end{eqnarray}
where $\Omega$ is an arbitrary function of $\psi$ and $ \tau,\alpha, \beta, \delta,\gamma, \theta, \sigma$
are arbitrary functions of $t$. Therefore, the symmetries depend on eight arbitrary functions.
\subsection{Nonclassical Lie symmetries}As it is well known, there exist the so called ``nonclassical symmetries" \cite{bluco}, \cite{olver} that are symmetries of the equation together with the ``invariant surface conditions"
\begin{eqnarray}
\phi_1=\xi_1u_x+\xi_2u_y+\xi_3u_t\nonumber\\
\phi_2=\xi_1v_x+\xi_2v_y+\xi_3v_t\label{2.4}\\
\phi_3=\xi_1\psi_x+\xi_2\psi_y+\xi_3\psi_t+\xi_4\psi_{\lambda}\nonumber
\end{eqnarray}
These conditions allow us to eliminate more derivatives of the fields in the determining equations. As is
well known, this elimination differs, depending on whether the values of $\xi_1, \xi_2, \xi_3$ are zero or
not. We  looked for these symmetries in (\ref{1.2}) but all of them are contained in (\ref{2.3}). Therefore,
the nonclassical method does not  provide new symmetries.

Let us now determine the $1+1$ spectral problems in $1+1$ dimensions derived from the different possible
reductions of (\ref{2.3}).

\section{Similarity  reductions of the  spectral problem}
There are several independent reductions depending on whether the arbitrary functions that appear in
(\ref{2.3}) are zero or not. We will classify the reductions in five classes.
\subsection*{I. Reductions for $\tau\neq 0$.}
We can solve the characteristic equation (\ref{2.2}) and  we find the following results
\begin{itemize}
\item Reduced variables:
The reduced variables $z_1$, $z_2$ can be defined as
\begin{equation}
 z_1=B_1y-B_2,\quad \quad z_2=B_1^2\tau x-C_1B_1^2y^2-B_1B_3y-B_4 \label{3.1}\end{equation}
\item Spectral parameter: Let $\Lambda$ be the reduced spectral parameter. Thus, it is obtained as:

\begin{equation}
\Lambda=\tau B_1\lambda+C_2B_1y+(\delta B_1+B_3) \label{3.2}\end{equation}
\item Reduced Fields:  \begin{eqnarray} && \int \frac{d \psi}{\Omega(\psi)}=e^{{ \int \frac{dt}{\tau(t)}}}\Phi(z_1,z_2,\Lambda)\nonumber\\
&& u(x,y,t)=\frac{F(z_1,z_2)+B_5}{\tau^2B_1^2}+\frac{C_2B_1x+N_1y^2+N_2y}{\tau B_1}\label{3.3}\\
&& v(x,y,t)=\frac{H(z_1,z_2)+B_6}{\tau^2B_1^3}+\left(\frac{N_3B_1y+N_4}{\tau B_1}\right)x\nonumber\\ &&
+\frac{N_5B_1^2y^3+N_6B_1y^2+N_7y}{\tau^2B_1^2}\nonumber\nonumber\end{eqnarray} where $F(z_1,z_2)$ and
$H(z_1,z_2)$ are the reduced fields and $\Phi(z_1,z_2)$ is the reduced eigenfunction.

The derivatives  of the functions $B_i=B_i(t),\, (i=1..6)$ are related to the seven arbitrary functions
$\tau, \alpha,\delta,\beta,\gamma,\theta,\sigma$. The explicit relations are shown in Appendix A.
$C_1=C_1(t),\, C_2=C_2 (t)$ and $N_i=N_i(t), (i=1..7)$ are auxiliary functions that are also explicitly
written  in Appendix A.

\item Reduced spectral problem:
We can now substitute the reductions (\ref{3.1})-(\ref{3.3}) in (\ref{1.2}) to obtain the following Lax pair
in $1+1$ dimensions:
\begin{eqnarray}&& \frac{\partial \Phi}{\partial z_1}+ \frac{\partial \Phi}{\partial z_2}\left(H_{z_2}+\Lambda\right)-
F_{z_2}\frac{\partial \Phi}{\partial
\Lambda}=0\nonumber\\
&& \left(F_{z_1}-\Lambda F_{z_2}\right)\frac{\partial \Phi}{\partial \Lambda}+ \left(F-H_{z_1}+\Lambda
H_{z_2}+\Lambda^2\right)\frac{\partial \Phi}{\partial z_2}+\Phi=0\label{3.4}\end{eqnarray}

\item Reduced Equations: The compatibility condition of (\ref{3.4}) yields the following system of equations in $1+1$ dimensions:
\begin{eqnarray}&&\left(FF_{z_2}-H_{z_1}F_{z_2}\right)_{z_2}+\left(F_{z_1}+H_{z_2}F_{z_2}\right)_{z_1}=0\nonumber \\
&&H_{z_1z_1}+FH_{z_2z_2}+H_{z_2}H_{z_1z_2}-H_{z_1}H_{z_2z_2}=0,\label{3.5}
\end{eqnarray}
which contains as a particular cases the equations
\begin{itemize}
\item a) $H=0$
$$F_{z_1z_1}+FF_{z_2z_2}=0$$
\item b) $F=0$
$$H_{z_1z_1}+H_{z_2}H_{z_1z_2}-H_{z_1}H_{z_2z_2}=0$$
\item c) $F=H_{z_1}$
$$\left(H_{z_1}+\frac{H_{z_2}^2}{2}\right)_{z_1z_1}=0$$
\end{itemize}

\end{itemize}

\subsection*{II. Reductions for $\tau=0$, $\alpha\neq 0$.}

In this case it is useful to define the function
$$K=K(y,t)=y+\frac{\delta}{\alpha}.$$

\begin{itemize}
\item Reduced variables: Let $z_1,z_2$ be the reduced variables. They can be obtained by solving (\ref{2.2}) as:
\begin{eqnarray}z_1&=&\int M_1(t)dt
\nonumber \\z_2&=&\frac{x-B_2}{K^2M_1}-\frac{B_1}{KM_1}+ln(K)-M_2.\label{3.6}\end{eqnarray}

\item Spectral parameter: The reduced spectral parameter $\Lambda$ is
\begin{equation}\Lambda=\frac{\lambda -B_3}{KM_1}-ln(K)+z_2-M_3.\label{3.7}\end{equation}
.

\item Reduced Fields:  The integration of (\ref{2.2}) yields the following reductions for the fields:
\begin{eqnarray} && \int \frac{d \psi}{\Omega(\psi)}=K^{\left(\frac{1}{\alpha}\right)}
\Phi(z_1,z_2,\Lambda) \nonumber\\
 &&u(x,y,t)=\frac{d\,B_3}{dt}K+B_4+\nonumber\\
&&\quad+K^2M_1^2\left[F(z_1,z_2)+\frac{1}{2}\left(z_2-ln(K)+1-B_0\right)^2+M_4\right]\nonumber\\
 &&v(x,y,t)=\label{3.8}\\
&&\quad-2\left[z_2+M_2-ln(K)\right]K^2M_1N_1+\frac{dN_1}{dt}K^2+N_2K+B_5+\nonumber
\\&&+K^3M_1^2\left[H(z_1,z_2)+\frac{3}{2}\left(z_2-ln(K)+\frac{4}{3}-\frac{2}{3}B_0\right)^2+M_5\right],\nonumber\end{eqnarray}
where $F(z_1,z_2)$ and $H(z_1,z_2)$ are the reduced fields and $\Phi(z_1,z_2)$ is the reduced eigenfunction.

Functions $B_i=B_i(t),i=0..5$ and $N_i=N_i(t),i=1..2$ are defined in terms of the six arbitrary functions
$\alpha,\beta,\delta,\gamma,\theta,\sigma$. Their explicit expressions appear in Appendix B. The five
 $M_i=M_i(t), (i=1..5)$ functions are, in principle, arbitrary but we have fixed them in the forms that appear
in Appendix B in order to have the simplest form for the spectral problem.
\item Reduced Spectral problem: In this case, The Lax pair reduces  to the following non-autonomous form;
\begin{eqnarray}&&\frac{\partial \Phi}{\partial z_1}+ \left(F-3H\right)\frac{\partial \Phi}{\partial z_2}+
\left(\Lambda^2-\Lambda+3F-3H\right)\frac{\partial \Phi}{\partial
\Lambda}-\nonumber\\ &&\quad-e^{-z_1}(\Lambda+z_2)\Phi =0\label{3.9}\\
&&\left(H_{z_2}-F_{z_2}\right)\frac{\partial \Phi}{\partial \Lambda}+
\left(H_{z_2}+\Lambda\right)\frac{\partial \Phi}{\partial z_2}+e^{-z_1}\Phi =0.\nonumber\end{eqnarray}

\item Reduced Equations: Although (\ref{3.9}) is non-autonomous, it yields the following autonomous system:
\begin{eqnarray}&&H_{z_1z_2}+(F-3H)H_{z_2z_2}+H_{z_2}^2+H_{z_2}+3(F-H)=0\label {3.10}\\
&&F_{z_1z_2}+(F-3H)F_{z_2z_2}+F_{z_2}^2-H_{z_2}+2F_{z_2}+3(F-H)=0\nonumber
\end{eqnarray}

\item When $F=H$, the system includes the equation
$$(H_{z_1}-2HH_{z_2}+H)_{z_2}+3H_{z_2}^2=0, $$ which can be understood as a modified Hunter-Saxton
equation\cite{beals}. In this particular case, the Lax pair (\ref{3.9}) can be written in the autonomous form
:
$$(H_{z_2}+\Lambda)\frac{\partial \Psi}{\partial z_2}+H_{z_2z_2}\Psi=0 $$
$$ \frac{\partial \Psi}{\partial z_1}-2H\frac{\partial \Psi}{\partial z_2}+\Lambda(\Lambda-1)\frac{\partial \Psi}{\partial \Lambda}-(H_{z_2}-\Lambda)\Psi=0$$
by means of the transformation $\Psi=(\textrm{ln}(\Phi))_{z_2}.$
\end{itemize}
\subsection*{III. Reductions for $\tau=\alpha=0$, $\delta\neq 0$.}

\begin{itemize}
\item Reduced variables: The integration of the characteristic system provides the reduced variables \begin{eqnarray}
z_1&=&\int M_1\,dt\nonumber \\
 z_2&=&\frac{1}{M_2}\left(x-\frac{\beta y^2+2\gamma y}{2\delta}\right).\label{3.11}\end{eqnarray}

\item Spectral parameter: The reduction of the spectral parameter is:
\begin{equation}\Lambda=\frac{\lambda\delta+B_1y-M_3}{M_2}.\label{3.12}\end{equation}

\item Reduced Fields: The reduced fields are: \begin{eqnarray} && \int \frac{d \psi}{\Omega(\psi)}=
e^{\left(\frac{y}{\delta}+M_4\right)}\Phi(z_1,z_2,\Lambda)\nonumber\\
&& u(x,y,t)=\frac{M_2^2}{\delta^2}F(z_1z_2)+\frac{-\frac{dB_1}{dt}y^2+2\theta y }{2\delta}+N_1\label{3.13}\\
&& v(x,y,t)=\frac{M_2^2}{\delta}H(z_1z_2)+\frac{B_4y^3+B_3y^2+\left(2M_2B_2z_2+\sigma\right) y
}{\delta}+N_2,\nonumber\end{eqnarray} where $F(z_1,z_2)$ and $H(z_1,z_2)$ are the reduced fields.
$\Phi(z_1,z_2)$ is the reduced eigenfunction and $\Lambda$ is the reduced spectral parameter.

Functions $B_i=B_i(t),(i=0..4)$ are defined in terms of the  arbitrary functions
$\beta,\delta,\gamma,\theta,\sigma$. Their explicit expressions appear in Appendix C. The
$M_i=M_i(t),(i=1..3)$, $N_i=N_i(t,z_2), (i=1..3)$ functions are in principle arbitrary, but we can fix them
in the forms that appear in Appendix C in order to have the simplest form for the spectral problem.
\item Reduced Spectral problem:
\begin{eqnarray}
&& \left(H_{z_2}+\Lambda\right)\frac{\partial \Phi}{\partial z_2}- F_{z_2}\frac{\partial \Phi}{\partial
\Lambda}+\Phi=0\nonumber\\ && \frac{\partial \Phi}{\partial z_1} +F\frac{\partial \Phi}{\partial
z_2}-\Lambda\Phi=0\end{eqnarray}

\item Reduced Equations: The compatibility condition  yields:
\begin{eqnarray}&&H_{z_1z_2}+FH_{z_2z_2}=0\nonumber\\
&&F_{z_1z_2}+FF_{z_2z_2}+F_{z_2}^2=0.\label{3.15}
\end{eqnarray}
\item It is interesting to notice that the equation for $F$ is the non-dispersive KdV equation
$$(F_{z_1}+FF_{z_2})_{z_2}=0.$$
By eliminating $F$ between the two equations (\ref{3.15}), For $H$ we obtain the equation
\begin{eqnarray}&&F=-\frac{H_{z_1z_2}}{H_{z_2z_2}}\nonumber\\
&&\left[\frac{1}{H_{z_2z_2}}\left(\frac{H_{z_1z_2}^2-H_{z_1z_1}H_{z_2z_2}}{H_{z_2z_2}}\right)_{z_2}\right]_{z_2}=0,\nonumber
\end{eqnarray}
which can be integrated twice with respect to $z_2$. It yields the generalized Monge-Ampere equation
$$H_{z_1z_2}^2-H_{z_1z_1}H_{z_2z_2}=a(z_1)H_{z_2}+b(z_1).$$
\end{itemize}

\subsection*{IV. Reductions for $\tau=\alpha=\delta=0$, $\beta\neq0$.}

\begin{itemize}
\item Reduced variables: The integration of the characteristic system allows us to write the reduced variables
as:
\begin{eqnarray}z_1&=&\int \beta(t)dt\nonumber \\z_2&=&y+\frac{\gamma(t)}{\beta(t)}.\label{3.16}\end{eqnarray}

\item Spectral parameter: The reduced spectral parameter $\Lambda $ is:
\begin{equation}\Lambda=\frac{\lambda z_2+x}{\beta z_2}.\label{3.17}\end{equation}

\item Reduced Fields: The reduced fields are:
\begin{eqnarray} && \int \frac{d \psi}{\Omega(\psi)}=e^{\frac{\lambda}{\beta}}\,\Phi(z_1,z_2,\Lambda)\nonumber\\
&& u(x,y,t)=\left(B_2-B_1\,y\right)\frac{\beta\, x}{z_2}+\beta^2 \,F(z_1,z_2) \\&&
v(x,y,t)=\left(-B_1\,y^2+\left(B_2-B_3\right)y+B_4\right)\frac{\beta \,x}{z_2} +\frac{x^2}{z_2}+\beta^2\,
H(z_1,z_2),\nonumber
\end{eqnarray}
where $F(z_1,z_2)$ and $H(z_1,z_2)$ are the reduced fields. $\Phi(z_1,z_2)$ is the reduced eigenfunction.

Functions $B_i=B_i(z_1),(i=0..4)$ are defined in terms of the four arbitrary functions
$\beta,\gamma,\theta,\sigma $. Their explicit expressions appear in Appendix D.
\item Reduced Spectral problem:
\begin{eqnarray}&& \frac{\partial \Phi}{\partial z_1}+\left(\frac{dB_0}{dz_1}-\Lambda\right)\frac{\partial \Phi}{\partial z_2}+
\left[F_{z_2}-\Lambda B_1+\frac{1}{z_2}\left(F-H_{z_2}\right)\right]\frac{\partial \Phi}{\partial
\Lambda}+\nonumber \\ && \quad +\left(F_{z_2} -\Lambda B_1\right)\Phi=0\nonumber\\
&& \frac{\partial \Phi}{\partial
z_2}+\frac{1}{z_2}\left(\Lambda-\frac{dB_0}{dz_1}+\frac{B_5}{z_2}\right)\frac{\partial \Phi}{\partial
\Lambda}+ \left(B_1-\frac{B_6}{z_2}\right)\Phi=0.\label{3.19}
\end{eqnarray}

\item Reduced Equations: The compatibility condition  yields the linear equations:
\begin{eqnarray}
&&\frac{d^2H}{dz_2^2}+\frac{2}{z_2}\left(F-\frac{d
H}{dz_2}\right)=\frac{B_5^2}{z_2^3}+\frac{B_6B_5}{z_2^2}-\frac{B_1B_5+\frac{dB_5}{dz_1}}{z_2}+\frac{d^2B_0}{dz_1^2}
+\nonumber \\
&& \quad+B_1\frac{dB_0}{dz_1}-\frac{dB_6}{dz_1}+z_2\frac{d B_1}{dz_1},\nonumber\\
&&\frac{d^2F}{dz_2^2}=\frac{B_5B_6}{z_2^3}-\frac{1}{z_2}\frac{dB_6}{dz_1}+\frac{dB_1}{dz_1},\nonumber
\end{eqnarray}
which can  easily be integrated as:
\begin{eqnarray}
&&H(z_1,z_2)=\frac{dB_6}{dz_1}\left(1-\textrm{ln}(z_2)\right)z_2^2+\frac{A_2(z_1)z_2^3}{2}+\nonumber\\
&&\quad
 +\frac{1}{2}\left(2A_1(z_1)-B_1\frac{dB_0}{
dz_1}-\frac{d^2B_0}{dz_1^2}\right)z_2^2+\label{3.20} \\
&&\quad
+\left(2C_1+\frac{dB_5}{dz_1}+B_1B_5\right)\frac{z_2}{2}+\frac{B_5^2}{4z_2}+C_2,\nonumber\\
&&F(z_1,z_2)=\frac{dB_6}{dz_1}z_2\left(1-\textrm{ln}(z_2)\right)+\frac{B_5B_6}{2z_2}
+\frac{z_2^2}{2}\frac{dB_1}{dz_1}+A_1z_2+C_1,\nonumber
\end{eqnarray}
where $A_1, A_2, C_1, C_2$ are arbitrary functions of $z_1$.
\end{itemize}
\subsection*{V. Reductions for $\tau=\alpha=\delta=\beta=0$, $\gamma\neq0$.}

\begin{itemize}
\item Reduced variables:
The reduced variables $z_1$ and $z_2$ are: \begin{eqnarray}z_1&=&\int
\gamma(t)dt\nonumber\\z_2&=&y.\label{3.21}\end{eqnarray}

\item Spectral parameter: The reduced spectral parameter $\Lambda$ is:
\begin{equation}\Lambda=\frac{\lambda}{\gamma}.\label{3.22}\end{equation}

\item Reduced Fields:  The reduction of the fields is: \begin{eqnarray} && \nonumber\int \frac{d \psi}{\Omega(\psi)}=e^{\frac{x}{\gamma}}\,\Phi(z_1,z_2,\Lambda)\\
&& u(x,y,t)=\gamma B_1\,x+\gamma^2F(z_1z_2)\label{3.23}\\
&& v(x,y,t)=\gamma (B_2\,y+B_3)x+\gamma^2H(z_1,z_2)\nonumber\end{eqnarray} where $F(z_1,z_2)$ and
$H(z_1,z_2)$ are the reduced fields and $\Phi(z_1,z_2),$ is the reduced eigenfunction.

Functions $B_i=B_i(z_1),(i=0..3)$ are defined in terms of the  arbitrary functions $\gamma,\theta,\sigma$.
Their explicit expressions appear in Appendix E. \item Reduced spectral problem:
\begin{eqnarray}&& \frac{\partial
\Phi}{\partial z_1} -\Lambda\frac{\partial \Phi}{\partial
z_2}+\left[F_{z_2}+\Lambda(B_2-B_1)\right]\frac{\partial \Phi}{\partial
\Lambda}+\left(F-H_{z_2}\right)\Phi=0\nonumber\\
&& \frac{\partial \Phi}{\partial z_2}- B_1\frac{\partial \Phi}{\partial
\Lambda}+\left(B_2z_2+B_3+\Lambda\right)\Phi=0,\label{3.24}\end{eqnarray}

\item Reduced Equations: The compatibility condition  yields the linear equations:
\begin{eqnarray}&&\frac{d^2F}{dz_2^2}=B_1B_2-2B_1^2-\frac{dB_1}{dz_1},\nonumber\\
&&\frac{d^2H}{dz_2^2}=-\left(B_1B_2+\frac{dB_2}{dz_1}\right)z_2-\left(B_1B_3+\frac{d
B_3}{dz_1}\right),\nonumber
\end{eqnarray}
which can  easily be integrated as:
\begin{eqnarray}&&F(z_1,z_2)=\left(B_1B_2-2B_1^2-\frac{dB_1}{dz_1}\right)\frac{z_2^2}{2}+A_1z_2+C_1,\label{3.25}\\
&&H(z_1,z_2)=\left(B_1B_2+\frac{dB_2}{dz_1}\right)\frac{z_2^3}{6}-\left(B_1B_3+\frac{dB_3}{dz_1}\right)\frac{z_2^2}{2}+\nonumber
\\ &&\quad +A_2z_2+C_2,\nonumber
\end{eqnarray}
where $A_1, A_2, C_1, C_2$ are arbitrary functions of $z_1$.
\end{itemize}

\section{Conclusions}
\begin{itemize}
\item We have studied the Lie symmetries of the spectral problem of the Manakov-Santini equation. The
procedure requires the consideration of the spectral parameter as an additional independent variable.
Therefore, the Lax pair would be considered as a system with three fields and four independent variables.
\item We have used computer packages such as MAPLE and MACSYMA to handle the calculation. The resulting
symmetries depend on seven arbitrary functions of $t$ and one arbitrary function of $\psi$.
\item We have also looked  for nonclassical symmetries and have realized that they are no different from
the classical ones.
\item Five independent reductions arise from the symmetries identified. The spectral problems  are obtained for all the
reductions.  Two of them give rise to nonlinear equations that can be easily integrated. The other three
yield reduced systems that include non dispersive KdV, generalized Monge-Ampere and modified Hunter-Saxton
equations, among others.

\end{itemize}

\section*{Acknowledgements}
This research has been supported in part by the DGICYT under project  FIS2009-07880 and JCyL under contract
GR224.

\newpage

\section*{Appendix A} The functions $B_i(t), C_i(t), N_i(t)$ that appear in equations (\ref{3.1})-(\ref{3.3})
are:

\begin{eqnarray}&&\frac{dB_1}{dt}=-\frac{\alpha B_1}{\tau}\nonumber\\
&&\frac{dB_2}{dt}=\frac{\delta B_1}{\tau}\nonumber\\
&&\frac{dB_3}{dt}=\beta B_1-2C_1\frac{dB_2}{dt}\nonumber\\
&&\frac{dB_4}{dt}=\gamma B_1^2-B_3\frac{dB_2}{dt}\nonumber \\
&&\frac{dB_5}{dt}=B_1^2\tau\theta-\delta B_1N_2-B_1^2c_2\gamma\nonumber \\
&&\frac{dB_6}{dt}=\sigma \tau B_1^3-N_7\frac{dB_2}{dt}-\gamma N_4B_1^2\nonumber \\&&\nonumber\\
&&C_1(t)=\frac{\tau_t}{2}-\alpha\nonumber\\&&C_2(t)=\tau_t-\alpha\nonumber\\&&\nonumber\\
&&N_1=-\frac{1}{2}\frac{d(C_2B_1)}{dt}\nonumber\\
&&N_2=-\frac{d(B_3+\delta B_1)}{dt}\nonumber\\
&&N_3=C_2+2C_1\nonumber\\
&&N_4=2B_3+\delta B_1\nonumber \\
&&N_5=-\frac{1}{6}\frac{d(\tau N_3)}{dt}+\frac{1}{2}\alpha N_3\nonumber \\
&&N_6=-\frac{1}{2}\frac{d(\tau N_4)}{dt}+\alpha N_4\nonumber \\
&&N_7=B_5-B_3^2-\tau\gamma B_1^2\nonumber
\end{eqnarray}
\section*{Appendix B}
$B_i(t), M_i(t), N_i(t)$ that appear In equations (\ref{3.6})-(\ref{3.8})  several functions $B_i(t), M_i(t),
N_i(t)$ of $t$ appear; They are are defined as:
\begin{eqnarray}&&B_0=\frac{1}{M_1^2}\frac{dM_1}{dt}\nonumber\\
&&B_1=\frac{-\beta-2\delta M_1}{\alpha}\nonumber\\
&&B_2=\frac{\delta^2M_1+\delta\beta-\gamma\alpha}{2\alpha^2}\nonumber\\
&&B_3=\frac{1}{\alpha}\left(\delta M_1+\beta+\frac{d\delta}{dt}\right)\nonumber \\
&&B_4=\frac{1}{4\alpha^2} \left(\beta\delta M_1-\alpha\gamma
M_1-\delta^2\frac{dM_1}{dt}-2\frac{d^2\delta}{dt^2}-2\delta\frac{d\beta}{dt}\right)
-\frac{\theta}{2\alpha}\nonumber \\
&&B_5=\frac{\delta}{6\alpha^3}\left(\delta^2\frac{dM_1}{dt}-2\delta^2M_1^2+2\delta\frac{d\beta}{dt}+
\delta\frac{d^2\delta}{dt^2}-5\beta\delta M_1-2\beta^2-\delta
M_1\frac{d\delta}{dt}-\beta\frac{d\delta}{dt}\right)+\nonumber\\ &&
+\frac{1}{6\alpha^2}\left(2\beta\gamma+2\delta\theta+3\delta\gamma
M_1+\gamma\frac{d\delta}{dt}-2\delta\frac{d\gamma}{dt}\right)-\frac{\sigma}{3\alpha}\nonumber
\\&&\nonumber\\&&N_1=-\frac{1}{2}\left(B_3-B_1\right)\nonumber\\
&&N_2=B_4-B_1B_3-\frac{dB_2}{dt}\nonumber
\\&&\nonumber\\&&M_1=\frac{1}{\alpha}\frac{d\alpha}{dt}\nonumber\\
&&M_2=\frac{3-B_0}{2}\nonumber\\
&&M_3=B_0-2\nonumber\\
&&M_4=\frac{1}{2}\left(\frac{1}{M_1}\frac{dB_0}{dt}-B_0\right)\nonumber \\
&&M_5=\frac{1}{3}\left(\frac{1}{M_1}\frac{dB_0}{dt}-B_0\right)+\frac{1}{6}\nonumber
\end{eqnarray}
\section*{Appendix C}
The functions $B_i(t)$, $M_i(t)$ and $N_i(t,z_2)$ that appear in  (3.11)-(3.13) are:

\begin{eqnarray}
&&B_1=\beta+\frac{d\delta}{dt}\nonumber\\
&&B_2=\beta+\frac{1}{2}\frac{d\delta}{dt}\nonumber\\
&&B_3=\frac{1}{2}\left(\theta-\frac{d\gamma}{dt}+2\frac{\gamma}{\delta}B_2\right)\nonumber \\
&&B_4=\frac{1}{3} \left(\frac{\beta B_2}{\delta}-\frac{dB_2}{dt}\right) \nonumber
\\&&\nonumber\\&&M_1=\frac{M_2}{\delta^2}\nonumber\\
&&M_2=e^{\left(-\int \frac{\beta}{\delta}dt\right)}\nonumber\\
&&\frac{dM_3}{dt}=\theta-\frac{\beta M_3+\beta\gamma+\gamma\frac{d\delta}{dt}}{\delta}\nonumber\\
&&\frac{dM_4}{dt}=\frac{\gamma+M_3}{\delta^2}\\ &&\nonumber\\
&&N_1=\frac{M_2}{\delta}B_1z_2+\frac{\sigma}{\delta}-\frac{\gamma^2}{\delta^2}\nonumber \\
&&N_2=\frac{M_2(\gamma-M_3)}{\delta}z_2\nonumber
\end{eqnarray}
\section*{Appendix D}
In (3.16)-(3.18)  several functions $B_i(z_1)$ are introduced. These reads:

\begin{eqnarray}
&&B_0=\left(\frac{\gamma}{\beta^2}\right)_{[t=t(z_1)]}\nonumber\\
&&B_1=\left(\frac{\beta_t}{\beta^2}\right)_{[t=t(z_1)]}\nonumber\\
&&B_2=\left(\frac{\theta}{\beta^2}\right)_{[t=t(z_1)]}\nonumber\\
&&B_3=
\left(\frac{\gamma_t}{\beta^2}\right)_{[t=t(z_1)]}\nonumber \\
&&B_4=\left(\frac{\sigma}{\beta^2}\right)_{[t=t(z_1)]}\nonumber\\
&&B_5=\left(B_4-B_0B_2+\frac{1}{\beta}\frac{dB_0}{dt}\right)_{[t=t(z_1)]}\nonumber\\
&&B_6=\left(B_2+B_1B_0\right)_{[t=t(z_1)]}\nonumber
\end{eqnarray}
where we have used $\frac{dB_i}{dt}=\beta\frac{dB_i}{dz_1}$ according to the definition of $z_1$ in
(\ref{3.16})
\section*{Appendix E}
The functions $B_i(z_1)$ introduced in (3.20)-(3.22) are:
\begin{eqnarray}
&&B_1=\left(\frac{\theta+\gamma_t}{\gamma^2}\right)_{[t=t(z_1)]}\nonumber\\
&&B_2=\left(\frac{\theta}{\gamma^2}\right)_{[t=t(z_1)]}\nonumber\\
&&B_3=\left(\frac{\sigma}{\gamma^2}\right)_{[t=t(z_1)]}\nonumber
\end{eqnarray}


\section*{References}

\begin{thebibliography}{00}
\bibitem{beals}  R. Beals, D. Sattinger and J. Smigielski, \textit{Inverse scattering
solutions of the Hunter-Saxton equation, }Applicable Analysis\textbf{\ 78, (}%
2001), 255-269.

\bibitem{bluco} G. W. Bluman and J. D. Cole, \textit{Similarity Methods for Differential Equations, }Springer Verlag,
(1974).

\bibitem{bodganov} L. V. Bodganov, \textit{On a class of reductions of the Manakov-Santini hierarchy connected with the
interpolation system, }Journal of Physics A: Math. and Gen. \textbf{43}, (2010), 115206 (11 pp).

\bibitem{chang} J. H. Chang and Y. T Chen, \textit{Hodograph solutions for the generalized dKP equation, }arXiv:
0904.4595v1, (2009), 259-261.

\bibitem{estevez93} P. G. Est\'{e}vez, \textit{The direct method and the singular\ manifold method for the Fitzhug-Naguno
equations,} Phys. Lett. \textbf{A171}, (1992), 259-261.

\bibitem{estevez95} P. G. Est\'{e}vez, \textit{A KdV Equation in 2+1 Dimensions: Painlev\'{e} Analysis, Solutions and
Similarity Reductions, }Studies in Appl. Math. \textbf{95}, (1995) , 73-113.



\bibitem{egp05} P. G. Est\'{e}vez, M. L. Gandarias and J. Prada, \textit{Symmetry Reductions of a Lax Pair , }Phys. Lett.
A \textbf{343}, (2005), 40-47.

\bibitem{ibragimov} N. Ibragimov, \textit{Elementary} \textit{Lie} \textit{group analysis and ordinary differential
equations, }John Wiley, Chichester, New York, (1999), ISBN 0-471-97430-7.

\bibitem{leg} M. Legare, \textit{Symmetry Reductions of the Lax Pair of the Four-Dimensional Euclidean Self-Dual
Yang-Mills Equations, }Journal of Nonlinear Math. Phys. \textbf{3}, (1996), 266-285.

\bibitem{santini} S. V. Manakov and P. M. Santini, \textit{A hierarchy of integrable partial differential equations in 2+1
dimensions, }Theor. and Math. Phys. \textbf{152}, (2007), 1004-1011.

\bibitem{olver} P. J. Olver, \textit{Applications of Lie Groups to Differential Equations, }Springer Verlag, (1999).

\bibitem{pavlov} M. V. Pavlov, J. H. Chang and Y. T. Chen, \textit{Integrability of the Manakov-Santini hierarchy, }arXiv:
0910.2400v1, (2009).

\bibitem{stephani} H. Stephani, \textit{Differential equations. Their solutions using symmetries, }edited by M. Mac Callum,
Cambridge University Press, (1989).



\end{thebibliography}
\end{document}